\documentclass{ws-procs9x6}

\def\iso#1#2{\hbox{$^{#1}$#2}}
\def\LiBe{\iso{7}{Li}$(p,n)$\iso{7}{Be}}
\def\etal{{\it et al.}}
\begin{document}

\title{Dark Matter Search with Direction Sensitive Scintillator}

\author{H.~SEKIYA\footnote{\uppercase{P}resent address:
 \uppercase{C}osmic \uppercase{R}ay
\uppercase{G}roup, \uppercase{D}epartment of \uppercase{P}hysics, 
\uppercase{G}raduate \uppercase{S}chool of \uppercase{S}cience,
 \uppercase{K}yoto \uppercase{U}niversity \uppercase{K}itashirakawa, 
 \uppercase{S}akyo,
 \uppercase{K}yoto 606-8502, \uppercase{J}apan}, M.~MINOWA, Y.~SHIMIZU, W.~SUGANUMA}

\address{Department of Physics, School of Science, University of Tokyo\\
7-3-1 Hongo Bunkyo-ku, Tokyo, 113-0033\\
E-mail: sekiya@icepp.s.u-tokyo.ac.jp}

\author{Y.~INOUE}

\address{International Center for Elementary Particle Physics,
University of Tokyo\\
7-3-1 Hongo Bunkyo-ku, Tokyo, 113-0033}

\maketitle

\abstracts
{A direction sensitive dark matter detector using organic crystals is
being developed.
It exploits the anisotropic scintillation efficiency of organic crystals
with respect to the direction of nuclear recoils relative to
crystallographic axes.
A variation of about 7\% was observed
in the scintillation efficiency of carbon recoils
in a stilbene crystal for recoil energies of 30\,keV to 1\,MeV.
We have performed a pilot experiment at Kamioka
to prove the feasibility of this method.}

\section{Introduction}
It is believed that the galactic halo is composed of WIMP dark matter.
WIMPs could be directly detected by measuring the nuclear recoils
produced by their elastic scattering off nuclei in detectors\cite{jungman}, and
the most convincing signature of the WIMPs appears in the directions
of nuclear recoils because the earth's velocity through
the galactic halo is large ($\sim230$km/s).
Hence, detectors sensitive to the direction of the recoil nucleus
have a great potential to identify WIMPs.

Now, we are developing organic crystalline scintillators as direction
sensitive dark matter detector\cite{sekiya}. In this paper, 
we report on anisotropic scintillation property of stilbene 
crystals and its application to the dark matter search.

\section{Direction Sensitive Scintillator}
It is known that scintillation efficiency of organic crystals to heavy 
charged particles depends on the direction of the particles
with respect to the crystallographic axes\cite{birks}.
This property could be applied to a direction sensitive
WIMPs detector\cite{belli}.

We adopted stilbene crystal scintillators,
because of its relatively high light yield (30\% of NaI),
and the anisotropy --- known to be about 20\%
for 6.5-MeV $\alpha$ particles\cite{heckmann}.
However, the recoil energy given by WIMPs is much lower.

Therefore, we measured the angle and energy dependence of 
the scintillation response of carbon recoils in a stilbene crystal
in the low energy region with
nuclear recoil events caused by elastic scattering of neutrons.
In order to obtain high statistics and wide neutron energy range,
two neutron sources, 
\LiBe\ and \iso{252}{Cf},
were employed.
The \LiBe\ source run was performed at 3.2-MV Pelletron accelerator
of the Research Laboratory for Nuclear Reactors
at Tokyo Institute of Technology.
Pulsed proton beam interacted with a thin lithium target,
and pulsed neutrons were produced by the \LiBe\ reaction.
Details of the experimental setup are given in Ref.\ \refcite{sekiya}.

\begin{figure}[t]
  \centerline{%
    \includegraphics[scale=.3]{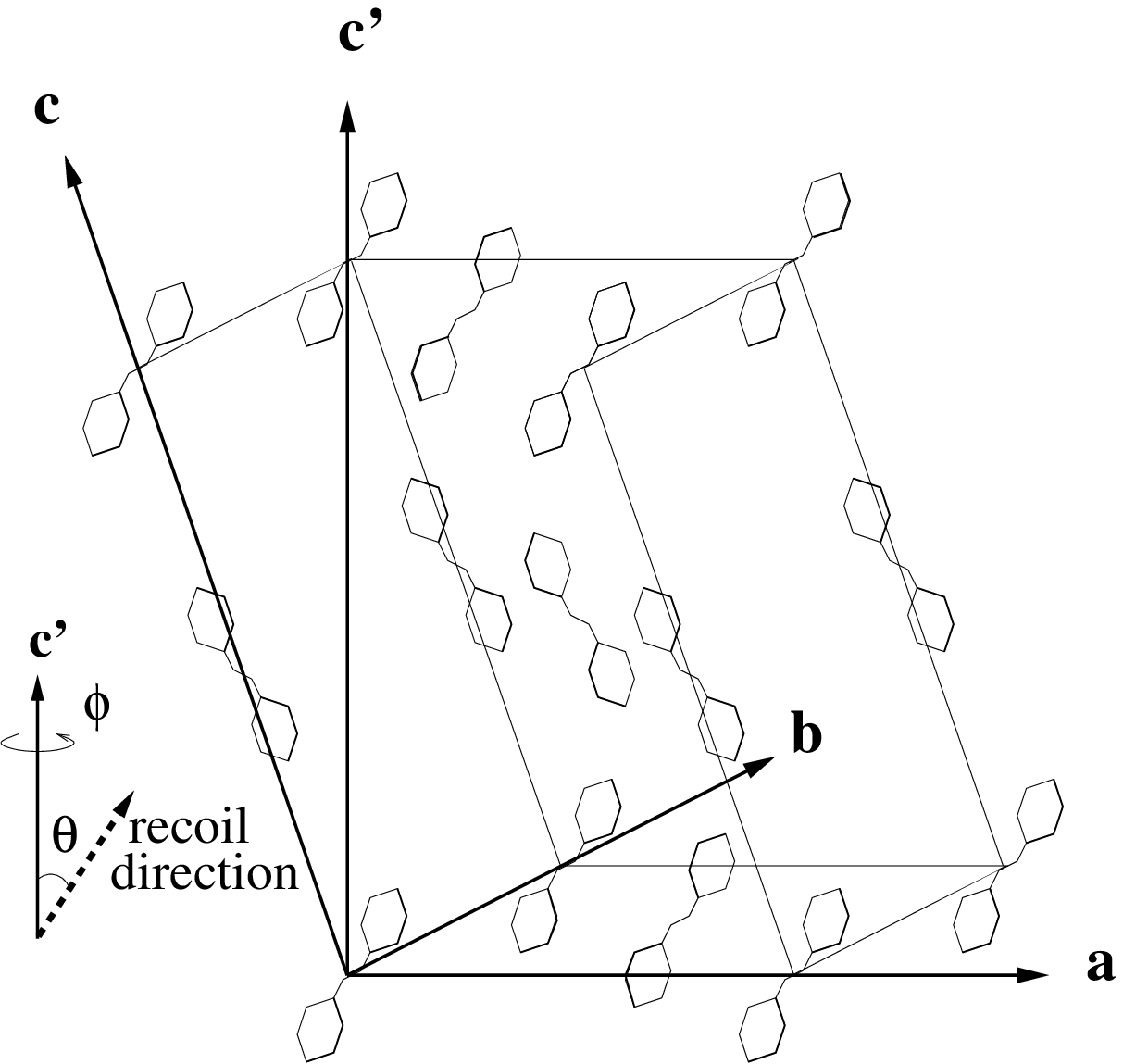}
    \hskip 1pc
    \includegraphics[scale=.6]{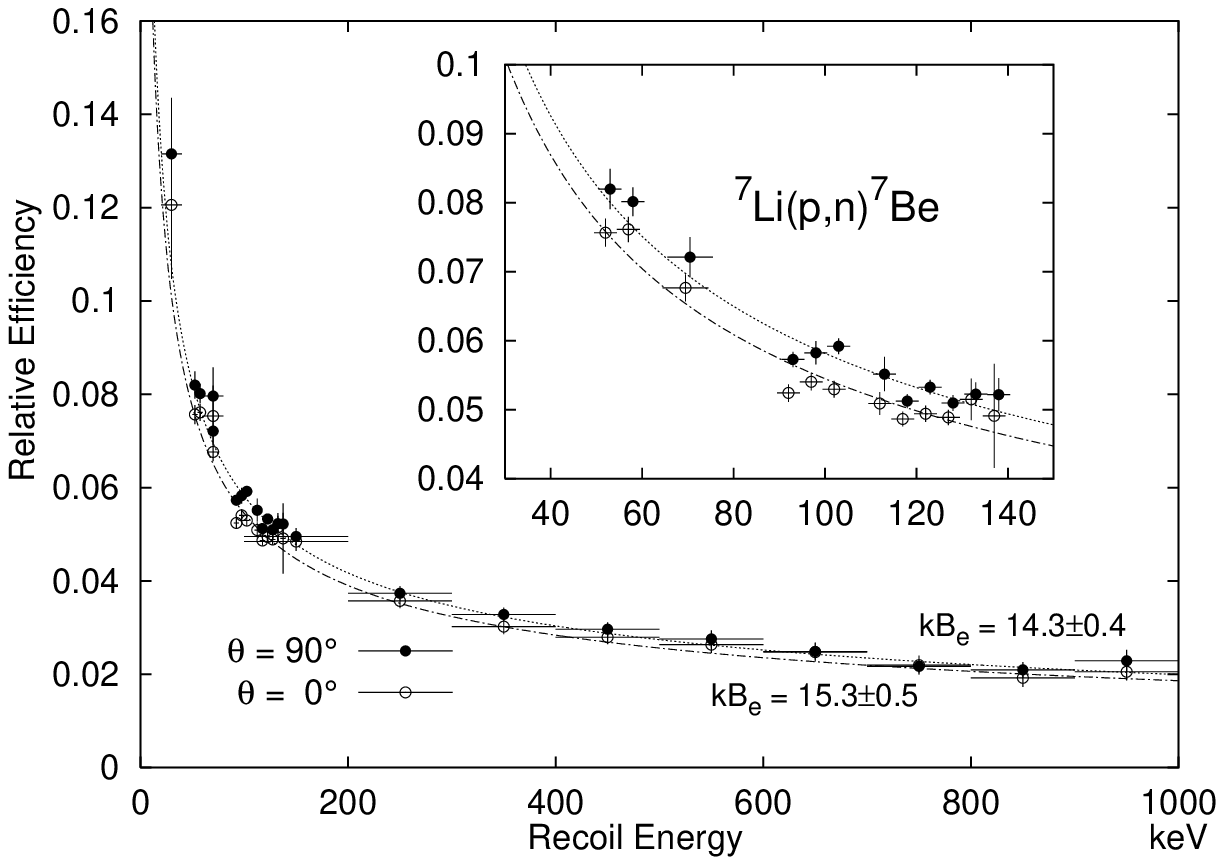}}
  \caption{
    The measured scintillation efficiency relative to that for
    electrons with $\theta=0^\circ$ and
    $90^\circ$ recoils ($\phi=0^\circ$).
    The inset is the results of the \LiBe\
    source run.
    Horizontal error bars represents selected recoil energy region
    for calculating the efficiency.
    The definition of the recoil angles $\theta$ and $\phi$
    are shown in the left along with the schematic drawing of the
    stilbene crystalline lattice
    where a stilbene molecule is drawn smaller than reality.
    $\theta$ is the angle with respect to the $c'$ axis
    and $\phi$ is the angle around the $c'$ axis.}
  \label{fig:anisotropy}
\end{figure}

Stilbene crystals form monoclinic systems
and the crystallographic axes are called $a$, $b$ and $c$
as shown in Fig.\ \ref{fig:anisotropy}.
The axis perpendicular to $a$--$b$ plane is called $c'$
which can be easily known since the crystal is cleaved
along $a$--$b$ plane.

A $2\times2\times2\rm\,cm^3$ stilbene crystal was used
as the target.
The energies of incident and scattered neutrons were measured by
the time-of-flight (TOF) method,
and the recoil energy $E_R$ was determined by kinematics.

The scintillation efficiency for carbon recoils
relative to that for electrons
with $\theta=0^\circ$ and $90^\circ$
are shown in Fig.\ \ref{fig:anisotropy}.
The variation of the scintillation efficiency in $\theta$ is about 7\%
over the measured energy region, $E_R=\textrm{30\,keV--1\,MeV}$.
The scintillation efficiency
is maximal in the direction perpendicular
to $c'$ axis namely at $\theta=90^\circ$,
and is minimal in the direction parallel to $c'$ axis
namely $\theta=0^\circ$.
No significant $\phi$-dependence is observed.
This behavior is consistent with
the result of proton recoils\cite{pikachu}
and of high energy charged particles\cite{heckmann}. 

\section{Application to WIMPs Detector}
\begin{figure}[t]
  \centerline{%
    \includegraphics[scale=.3]{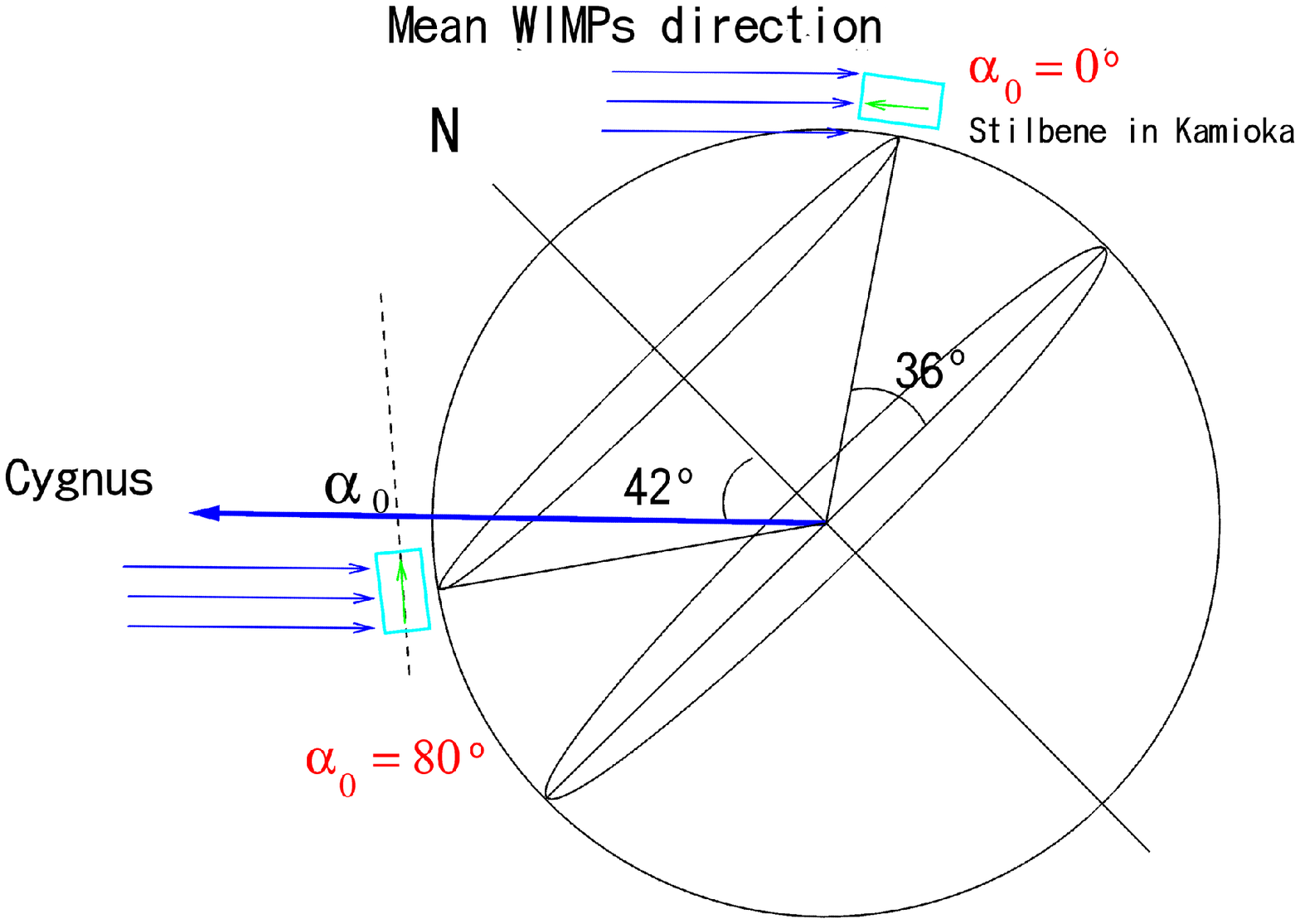}
    \hskip 1pc
    \includegraphics[scale=.5]{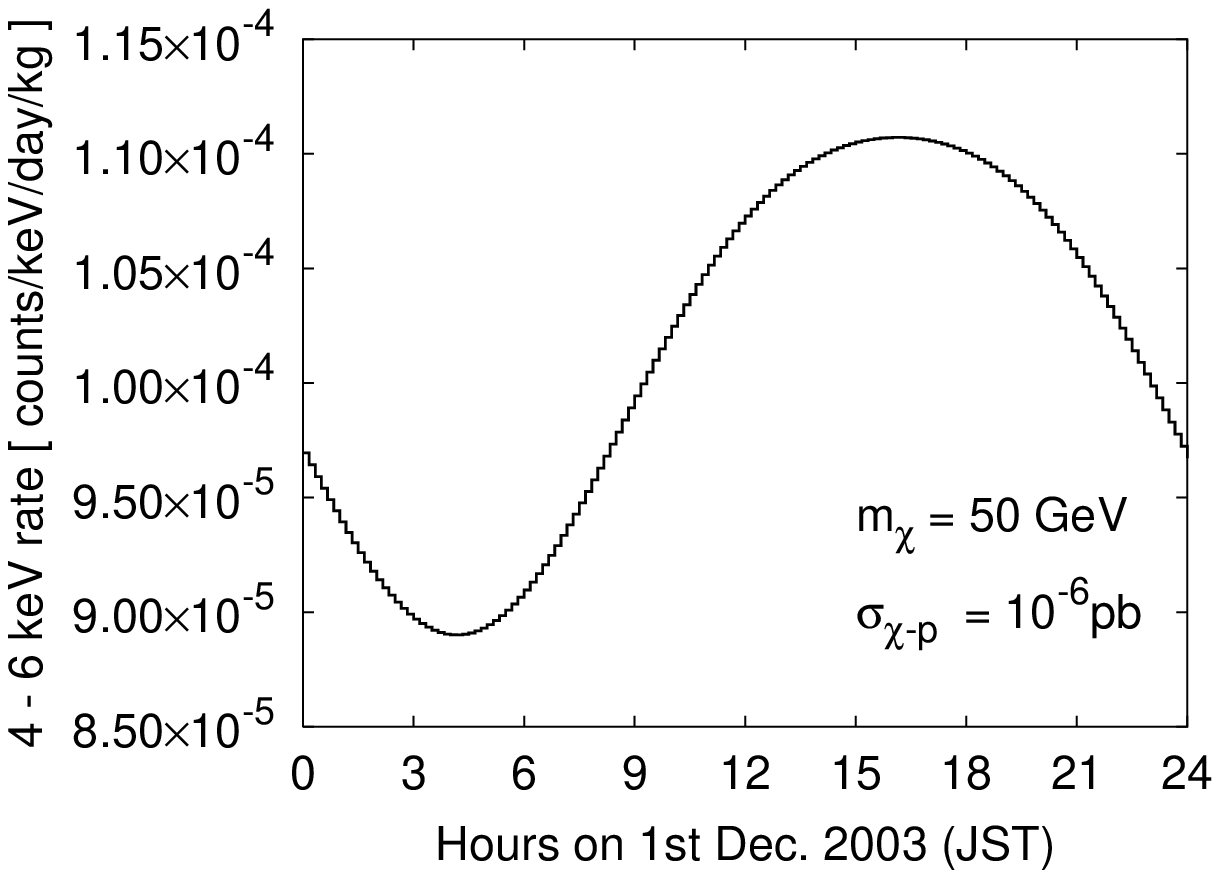}}
  \caption{left: Schematic drawing of the experimental approach method mentioned in
   the text. right: The expected signal variation of the counting rate
 of WIMP induced carbon recoil in
   an stilbene crystal with  $c'$ axis parallel to north-south direction 
   as of December 1st 2003 (JST). The  parameters that we used in the calculation are 
 $\rho_0=0.3$ GeV/cm$^3$, $v_0=220$ km/sec, 
 WIMP-proton spin independent cross section $\sigma_{\chi-p}=10^{-6}$ pb,
 and $m_{\chi}=50$ GeV}
  \label{fig:stilrotate}
\end{figure}

Let us discuss the directionality signature by considering the
expected signal counting rate. Assuming the WIMP halo is an 
isothermal sphere, the WIMPs mainly come 
from the direction of the Cygnus whose polar zenith angle 
is about 42$^{\circ}$ in the heliocentric frame.
If we perform the experiment in Kamioka Observatory at
36$^\circ$ 25' 30'' N and 137$^\circ$ 18' 32''E,
a suitable arrangement for the stilbene crystal is to
install the detector with the $c'$ axis in parallel to
the horizontal plane and towards the North as illustrated 
in Fig.\ \ref{fig:stilrotate}.
In that case, the mean incident angle of the WIMP  with respect 
to $c'$ axis, $\alpha_0$,  varies between $5^\circ$ and $78^\circ$
within a sidereal daily period neglecting the effect of 
aberration of WIMPs for simplicity.

An example of the variation of the
expected counting rate in 4-6 keV energy window calculated by 
the Monte Carlo method as of December 1st 2003 JST 
is shown in Fig.\ \ref{fig:stilrotate}. 
The effect of the motion of the earth is fully taken into account
using NOVAS\cite{NOVAS}.

\section{Pilot Experiment in Kamioka Observatory}
\begin{figure}[t]
\begin{center}
    \includegraphics[scale=.4]{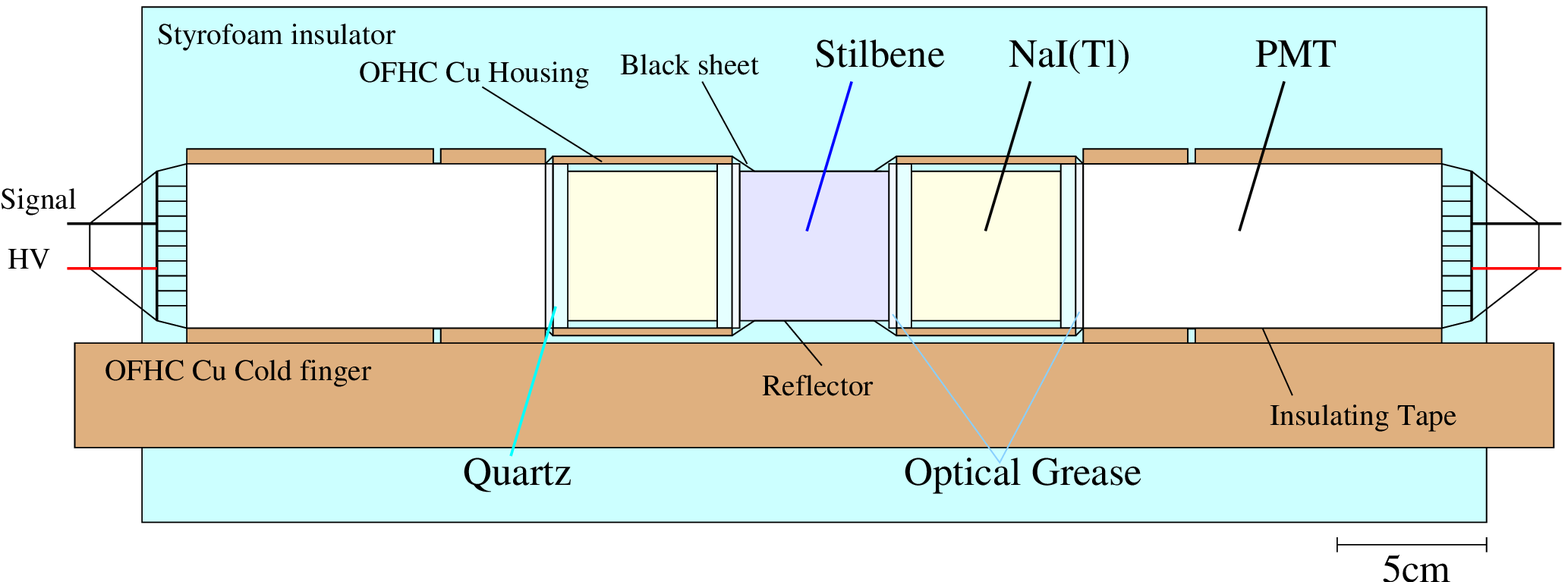}\\
    \includegraphics[scale=.5]{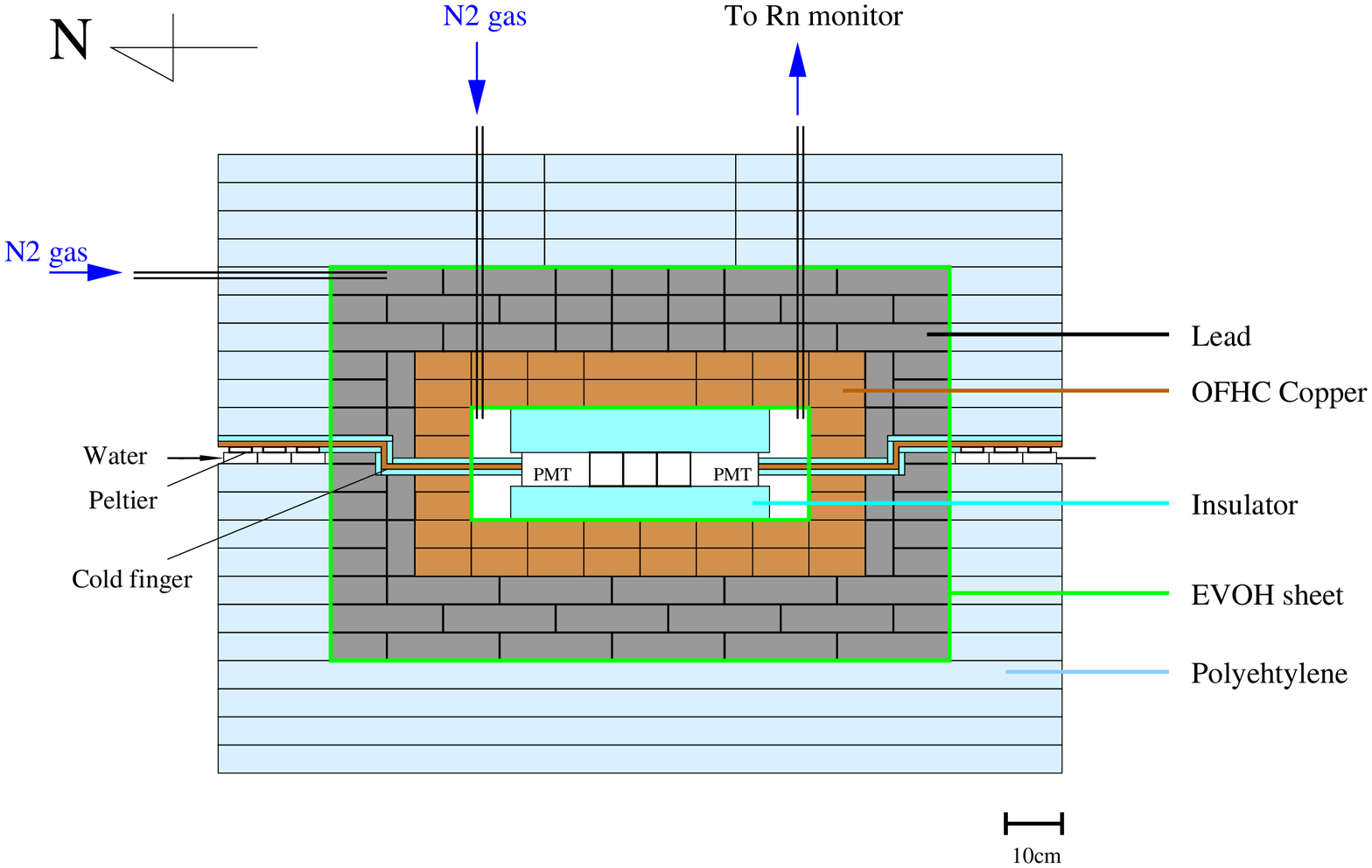}
\end{center}
  \caption{Schematic view of the detector and the shields.}
  \label{fig:pilot}
\end{figure}

A pilot dark matter search experiment to prove the feasibility 
of the method was carried out in Kamioka Observatory.
The schematic view of the experimental setup is shown
in Fig.\ \ref{fig:pilot}.
The $\phi50{\rm\,mm}\times50{\rm\,mm}$ (116\,g)
cylindrical stilbene crystal is viewed by two
Hamamatsu R8778MOD low background PMTs
through two Horiba low background NaI(Tl) active shields.
Self coincidence of two PMTs are required 
and both PMTs are cooled down at about $-7^\circ\rm\!C$ to 
reduce dark current further.
The detector assembly is shielded with 10cm OFHC copper, 15 cm Lead, and 
20cm polyethylene. The EVOH sheets\cite{evoh} are formed into
 air tight bags filled with nitrogen gas for purging the radon gas.
As mentioned above, the whole setup is laid with the
$c'$ axis of the crystal parallel to the north-south direction.

\begin{figure}[t]
\begin{center}
    \includegraphics[scale=.4]{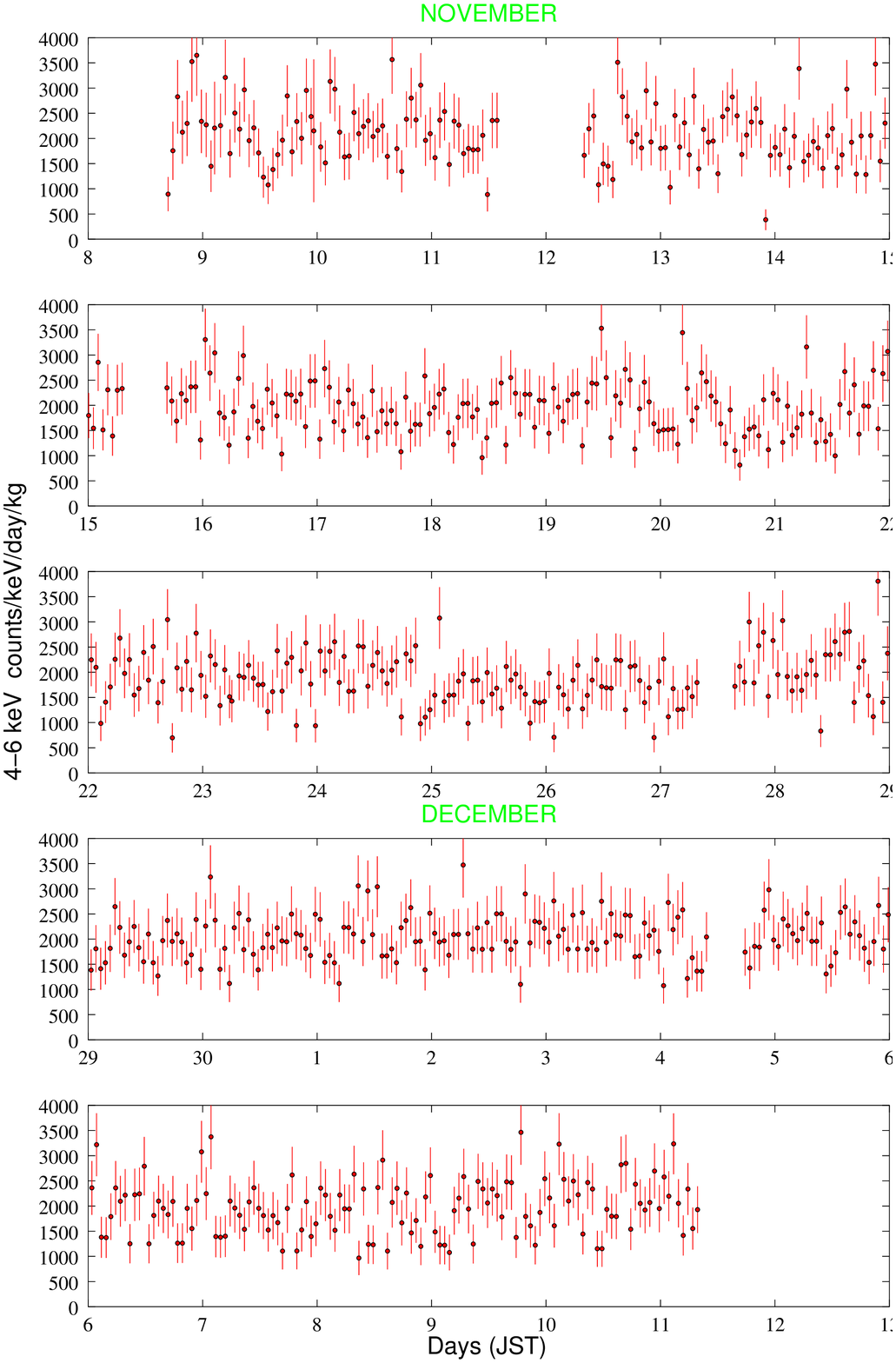}
  \caption{The obtained event rates for 4-6 keV during the measurements. }
  \label{fig:rate46}
\end{center}
\end{figure}

With the detector system, we started the measurement in October 25, 2003
and it was halted in December 11, 2003. The obtained count rate of
4-6 keV energy window for every one hour during the measurements
are shown in Fig.\ \ref{fig:rate46}.
Since the background level is as high as 2000 counts/keV/day/kg,
no modulation signal was found in the results.

\section{Discussions and Future Prospects}
It is obvious that rather high background rate due to the radioactivity
in  PMTs limits the sensitivity. Although the introduction of NaI(Tl) was
turned out to be somewhat effective, it is not sufficient to deal with
the problem. Therefore, in order to overcome the difficulties, highly radio-pure,
high quantum efficiency, high gain photon detector
is indispensable. Recently, the technology to produce avalanche photodiodes
(APDs) has improved significantly, and comparable or better performances
of APDs to those of PMTs are obtained with various scintillators\cite{APD}.
Hence, we are going to develop APDs with low background materials. 

Another disadvantage of the current method is the atomic (i.e.$^{12}$C) weights.
Generally, the SI cross section is proportional to $A^2$. 
On the other hand, the SD cross section is proportional to the nuclear
spin factor $\lambda^{2}J(J+1)$ and ${\rm ^{19}F}$ is considered to be 
the best nucleus to search for the SD interaction\cite{bolo}.
Therefore, we focus on fluorine loaded organic crystals.
We are now processing the growth of single crystals of 
octafluolonaphthalene (C$_{10}$F$_{8}$).

\end{document}